\documentclass[letterpaper,conference]{IEEEtran}
\IEEEoverridecommandlockouts

\usepackage[english]{babel}
\usepackage{cite}
\usepackage{amsmath,amsfonts,amssymb}

\DeclareFontFamily{U}{rcjhbltx}{}
\DeclareFontShape{U}{rcjhbltx}{m}{n}{<->rcjhbltx}{}
\DeclareSymbolFont{hebrewletters}{U}{rcjhbltx}{m}{n}

\DeclareMathSymbol{\tav}{\mathord}{hebrewletters}{116}
\DeclareMathSymbol{\tsadi}{\mathord}{hebrewletters}{118}
\DeclareMathSymbol{\samekh}{\mathord}{hebrewletters}{115}
\DeclareMathSymbol{\bet}{\mathord}{hebrewletters}{98}
\usepackage{graphicx}
\usepackage{caption}
\usepackage{subcaption}
\usepackage{textcomp}
\usepackage[colorlinks=true, allcolors=black]{hyperref}
\usepackage{xcolor}

\usepackage{algorithm}

\usepackage{algpseudocode}

\usepackage{microtype}

\begin{document}

\title{\vspace{-2ex}Memristor-based hardware and algorithms for higher-order Hopfield optimization solver outperforming quadratic Ising machines}

\author{
	\\[-2em]
	\IEEEauthorblockA{Mohammad~Hizzani\IEEEauthorrefmark{1}\IEEEauthorrefmark{2}, Arne~Heittmann\IEEEauthorrefmark{1}, George~Hutchinson\IEEEauthorrefmark{3}, Dmitrii~Dobrynin\IEEEauthorrefmark{1}\IEEEauthorrefmark{2},\\ Thomas~Van~Vaerenbergh\IEEEauthorrefmark{4}, Tinish~Bhattacharya\IEEEauthorrefmark{3}, Adrien~Renaudineau\IEEEauthorrefmark{5}, Dmitri~Strukov\IEEEauthorrefmark{3}, John~Paul~Strachan\IEEEauthorrefmark{1}\IEEEauthorrefmark{2}}

\IEEEauthorblockA{\IEEEauthorrefmark{1}Forschungszentrum Jülich GmbH, Jülich, Germany}
\IEEEauthorblockA{\IEEEauthorrefmark{2}RWTH~Aachen University, Aachen, Germany}
\IEEEauthorblockA{\IEEEauthorrefmark{3}University of California, Santa Barbara, Santa Barbara, CA, USA}
\IEEEauthorblockA{\IEEEauthorrefmark{4}Hewlett Packard Enterprise, Brussels, Belgium}
\IEEEauthorblockA{\IEEEauthorrefmark{5}Université Paris-Saclay, Paris, France}

\IEEEauthorblockA{Corresponding Email: m.hizznai@fz-juelich.de}
\\[-3em]
}

\maketitle

\begin{abstract}

 Ising solvers offer a promising physics-based approach to tackle the challenging class of combinatorial optimization problems. However, typical solvers operate in a quadratic energy space, having only pair-wise coupling elements which already dominate area and energy. We show that such quadratization can cause severe problems: increased dimensionality, a rugged search landscape, and misalignment with the original objective function. Here, we design and quantify a higher-order Hopfield optimization solver, with 28nm CMOS technology and memristive couplings for lower area and energy computations. We combine algorithmic and circuit analysis to show quantitative advantages over quadratic Ising Machines (IM)s, yielding 48x and 72x reduction in time-to-solution (TTS) and energy-to-solution (ETS) respectively for Boolean satisfiability problems of 150 variables, with favorable scaling.
\end{abstract}

\begin{IEEEkeywords}
Optimization, Hopfield neural network, Ising machine, Boolean satisfiability
\end{IEEEkeywords}
\vspace{-2ex}
\section{Introduction}

With diminishing performance gains from CMOS in traditional hardware, alternative mechanisms to deliver high-performance, particularly for challenging (NP-hard) optimization problems has growing importance. A variety of physical systems designed to accelerate intractable combinatorial optimization has attracted recent attention, in many cases implementing a quadratic Ising solver in a novel computing substrate such as superconductor-based quantum annealing, optics, CMOS-based coupled oscillators, etc \cite{mohseni2022ising,Schuetz2022, goto2021high, mazyavkina2021reinforcement,opticalising,optising,cimandquantann,aramonPhysicsInspiredOptimizationQuadratic2019}.

An Ising-based solver typically maps the desired optimization problem into a quadratic energy function to be minimized. Quadratic terms represent pair-wise interactions, which are typically easy to implement in a physical system, such as with resistive or capacitive couplings, while higher-order (3-body or higher) are more challenging. Numerous approaches exist for conversion of an arbitrary target objective function into quadratic form (so called QUBO: Quadratic Unconstrained Binary Optimization). However, it is shown here that a quadratic-only solver is severely disadvantaged compared to a solver supporting higher-order interactions (so called PUBO: Polynomial Unconstrained Binary Optimization). For example, there are naturally cubic terms in the case of 3-SAT problems. 
Quadratization requires the introduction of additional auxiliary variables, typically much more than the original problem variables, leading to an exponentially increased search space.  The new search space may also have additional ruggedness that hinders a solver, as will be illustrated later.

In the present work, we utilized 3-SAT optimization problems as representative for our proposed PUBO implementation. We chose 3-SAT because it is NP-Complete and has widespread applications \cite{3sateda,3satairtrafic,3satcrypanalysis}.

This paper is structured as follows: Section~\ref{sec:background} gives a brief introduction to Hopfield Neural Networks and 3-SAT. Section~\ref{sec:algcomp} compares QUBO and PUBO algorithms and highlights the advantages of the latter through novel energy landscape analysis. Finally, in Section~\ref{sec:hwmodels}, we present hardware circuit designs supporting both algorithms, enabling quantitative  comparisons of key metrics TTS and ETS.

\section{Background and Motivation}\label{sec:background}

Hopfield Neural Networks (HNNs), originally proposed by J. J. Hopfield in 1982 \cite{hnn1982}, are a type of recurrent neural network, the dynamics of which is governed by an overall energy function, as in Eq.~\ref{equ:hnn}. Just as in physical systems, the network evolves to minimize this energy, leading to a neuron update rule proportional to the negative energy gradient  with respect to each dynamical neuron.
\begin{equation}
    E(\{ s\}) = - \frac{1}{2} \sum_{\langle ij\rangle} w_{ij} s_i s_j + \sum_i b_i s_i,\ s_i\in \{-1, 1\}
    \label{equ:hnn}
\end{equation}

Ising Machines (IM) operate based on a Hamiltonian, a mathematical description of the energy of a physical system. In the context of IMs, the Hamiltonian represents the problem to be solved. It consists of two key components: the "spins" representing the variables in the problem, and their interactions which are quantified as coupling strengths (manifesting similar to HNN eq.~\ref{equ:hnn}). By manipulating these spins, IMs seek the lowest energy state of the system, which corresponds to the optimal solution for the given problem.

Satisfiability problems are typically represented in conjunctive normal form (CNF) as shown in eq.~\ref{equ:cnf}. The goal is to find input values that satisfy the Boolean function, making it \texttt{true}. In $k$-SAT, 'k' represents the maximum number of literals in a SAT clause (eq.~\ref{equ:cnf}). Each clause, such as $(x_i \vee \neg x_j \vee x_k)$, consists of positive $x_i$ and negative $\neg x_j$ literals.
\begin{equation}
F(x_1, \dots, x_N) = (x_i \vee \neg x_j \vee x_k) \wedge (x_a \vee x_b \vee \neg x_c) \wedge \dots
\label{equ:cnf}
\end{equation}

SAT solvers can be exact or non-exact. Exact solvers, like DPLL and CDCL \cite{dpll,dpll1, Bayardo1997, GRASP, Chaff}, verify SAT problem satisfiability or unsatisfiability. Non-exact solvers, such as SLS algorithms, use heuristics and random assignments for local search, including variable flips. Examples include QUBO IMs/HNNs, PUBO HNNs, and others relying on the \texttt{make} and \texttt{break} evaluation such as WalkSAT, GSAT, and probSAT \cite{walksat, probSAT}.

\begin{figure}
    \centering

    \includegraphics[width=\columnwidth]{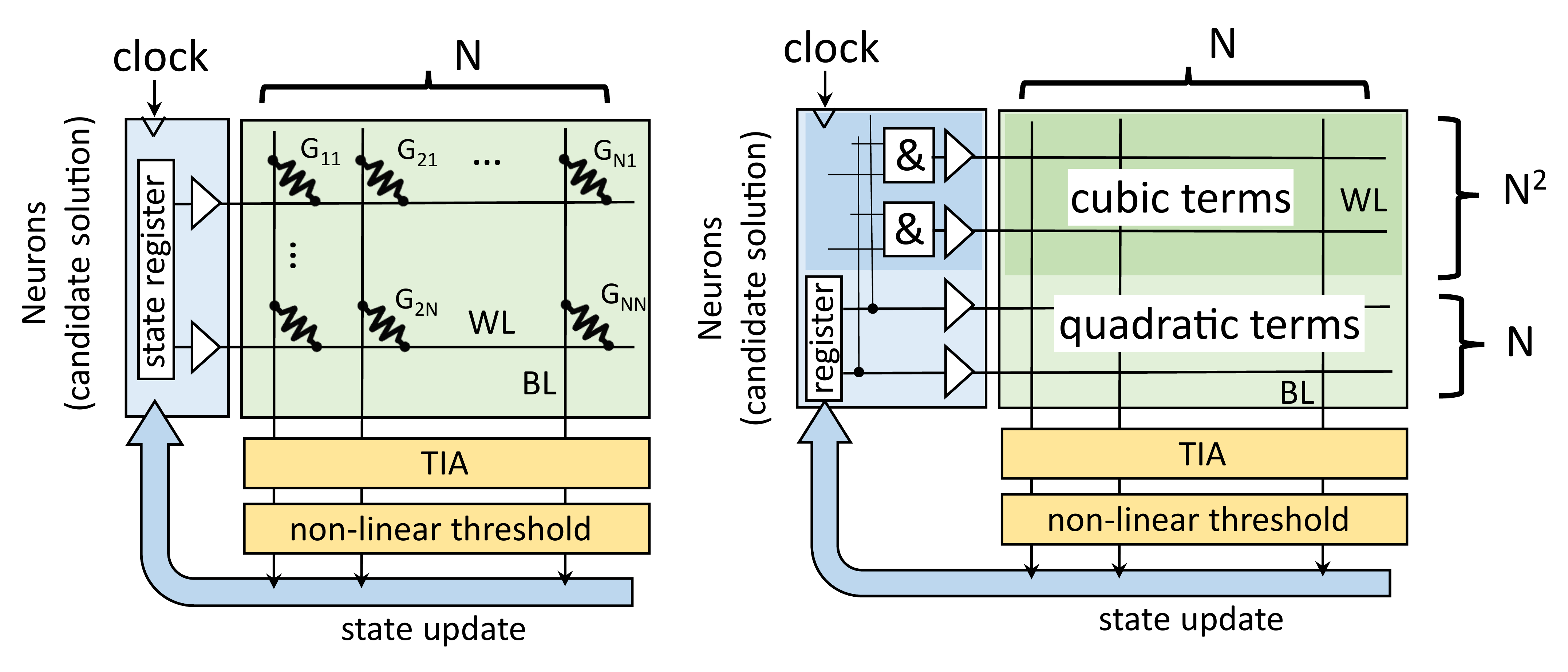}
    \caption{Left: QUBO HNN that utilizes memristors to store synapse weights in conductance to perform VMM. Right: PUBO HNN with support for cubic interactions. N is number of variables (n\textsubscript{var})}
    \label{fig:qubocartoon}
    \vspace{-1.0ex}
\end{figure}

Recent advancements in non-volatile analog memory technologies\cite{wang2020resistive, sebastian2020memory}, particularly 2-terminal memristive devices\cite{yang2013memristive, rao2023thousands}, have made mixed-signal implementations of certain computing primitives increasingly attractive due to their low area footprint, monolithic 3D potential, and multi-bit capacity. Furthermore, a cross-bar array of memristive devices can store coupling matrices and efficiently perform vector-matrix multiplications (VMM), as needed for both QUBO (2D matrix) and PUBO (3D tensor), allowing for computing-in-memory (CIM), as illustrated in Fig.~\ref{fig:qubocartoon}.

\vspace{-1ex}
\section{Comparing quadratic and higher-order solvers and energy landscapes}\label{sec:algcomp}

This section compares quadratic (QUBO) and higher-order (PUBO) solvers in terms of optimized algorithms, the solution space that must be navigated, and introduces some novel techniques for visualization. Performance is compared quantitatively, and 3-SAT problems are used for concrete analysis . The development and testing of algorithms necessitates adaptability while considering hardware compatibility. To ensure a fair comparison, both algorithms underwent in-depth optimization.

\vspace{-2ex}
\subsection{QUBO}

Quadratization of higher-order problems, such as 3-SAT, typically involve substituting products of variables $x_i, x_j$ with a new auxiliary variable $y$ and introducing a constraint to maintain this substitution, converted into a penalty term added to the objective function (Alg.~\ref{alg:qubo}). 
A penalty term, initially introduced by Rosenberg \cite{Rosenberg1975}, emerged as highly effective, especially when combined with stochastic group parallel updates. In essence, a QUBO HNN is governed by an energy function:
\begin{equation}
    \small
    E(\{s\}) = \sum_{\langle ij\rangle} w_{ij} s_i s_j + \sum_i b_i s_i + c ;
    \begin{cases}
    \{s\} = \{x\} \cup \{y\}\\
    s \in \{0, 1\}    
    \end{cases} 
\end{equation}
This function comprises quadratic terms, linear terms (biases), and a constant. While the resulting QUBO mapping is indeed quadratic, it exhibits high sparsity since the auxiliary variables ($y$) only couple to the original variables ($x$). This sparsity provides an opportunity for parallel updates within the algorithm, employing simulated annealing to overcome barriers.

\begin{algorithm}[!h]
	
	\footnotesize
	\caption{Quadratizing a 3-SAT with a Rosenberg penalty}
	\label{alg:qubo}
	\begin{algorithmic}
		\Require 3-SAT problem in CNF\\ //example $(\neg x_1 \vee x_2 \vee x_3) \wedge (x_1 \vee x_3 \vee x_4)$\\
		\textbf{Convert} dissafisfiability $\leftarrow$ Safisfiability (negating)\\ 
		//example $( x_1 \wedge\neg x_2 \wedge\neg x_3) \vee  (\neg x_1 \wedge\neg x_3 \wedge\neg x_4)$\\
		\textbf{Unroll} each clause into penalty\\ 
		//example $(x_1 (1-x_2)(1-x_3)) + ((1-x_1)  (1-x_3) (1-x_4))$\\
		\textbf{Substitute} the product of two variables in cubic term with aux variable\\ 
		//example $(y_1 (1-x_3)) + (y_2 (1-x_4)) : x_1 (1-x_2) = y_1, (1-x_1)  (1-x_3) = y_2$\\
		\textbf{Add} penalty for each auxiliary equality constraint\\
		//example $(y_1 (1-x_3)) + (y_2 (1-x_4)) + P(x_1(1-x_2) - 2x_1 y_1 -2 (1-x_2)y_1 + 3y_1) + P((1-x_1)(1-x_3) - 2(1-x_1)y_2 - 2(1-x_3)y_2 + 3y_2) : P $ is a parameter
	\end{algorithmic}
\end{algorithm}
\vspace{-1ex}

To implement stochastic parallel updates, we randomly distribute neurons into a desired number of groups at each step, then update neurons of each group sequentially. This follows an annealing schedule starting at high temperature to zero. We found stochastic parallel updates significantly improve convergence time of QUBO HNN for 3-SAT compared with all single-neuron update algorithms tested. 

A detailed analysis of the QUBO HNN's behavior revealed cases where the solver traversed through a satisfiability condition (i.e. 3-SAT problem solved) despite the QUBO energy not reaching zero, leading to a departure from the solution. To address this challenge, we introduced an additional component, the SAT checker, to monitor the solver's evolution cycle and halt the search upon finding satisfiability. This enhancement significantly accelerated the QUBO HNN (Fig.~\ref{fig:ttscycle}a), which we implemented using a cross-bar array with variables at word-line and clauses at bit-line, the clause is satisfied when the current is above a threshold representing zero.

\vspace{-2ex}
\subsection{PUBO}
To natively accommodate the energy function of 3-SAT within the Hopfield Neural Network (HNN), cubic interactions must be supported. The conversion of 3-SAT into this function closely resembles Alg.~\ref{alg:qubo}, albeit without variable substitution, resulting in the following equation:
\begin{equation}
	\scriptsize
    E(\{s\}) =\sum_{\langle ijk\rangle} ijkw_{ijk} s_i s_j s_k +\sum_{\langle ij \rangle} w_{ij} s_i s_j + \sum_i b_i s_i + c ;\ s \in \{0, 1\} 
\end{equation}%

Adhering to the classical HNN update rule of one neuron per step was observed to give quickest convergence and superior scaling compared to our best QUBO (see Fig.~\ref{fig:ttscycle}a). However, top performance was found by implementing a modified update rule inspired by Aramon, et al. \cite{aramonPhysicsInspiredOptimizationQuadratic2019} that increases the probability of flips. This starts by computing state changes for all spins, selects a flip if it exists (focus), if not it adds $\Delta E += E_\text{offset}$ to all gradients to induce a flip on  the next step (offset), and if a flip is found set $\Delta E = 0$.

\subsection{Landscape Advantages of PUBO over QUBO}\label{subsec:algadv}
\begin{figure}[!h]
 \vspace{-1em}
	\centering
	\includegraphics[width=1\columnwidth]{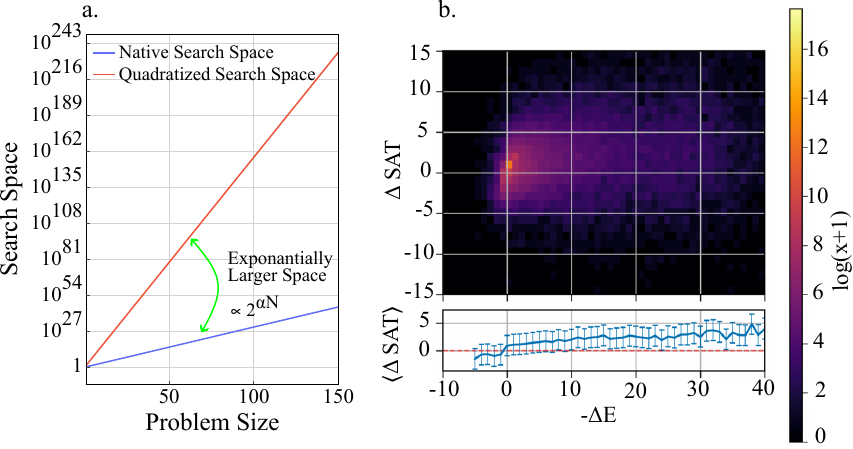}
	\caption{a. Search space vs problem size for quadratized and native space for hard 3-SAT random instances with number of clauses $4.23 \times$ number of variables, resulting in $2^N$ and $2^{5.23}$ size of search space for native and quadratized representation respectively. b. Histogram of running 3-SAT QUBO solver, showing energy reduction ($-\Delta E$) does not correlate well with satisfying more clauses $\Delta \text{SAT}$ .}
	\label{fig:qubobspubosearchspace}
     \vspace{-1.5ex}
\end{figure}

\begin{figure}[!h]
    \centering
    \includegraphics[width=\columnwidth]{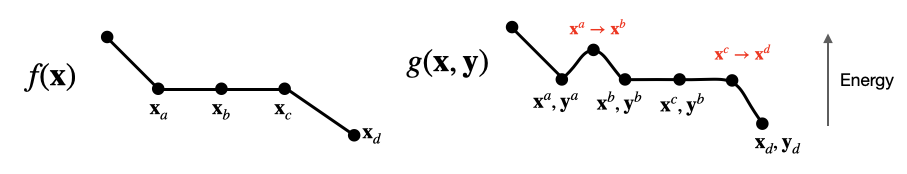}
    \caption{The deformation of the native (PUBO) energy landscape $f(\mathbf{x})$
    when reformulated as a QUBO problem $g(\mathbf{x}, \mathbf{y})$ with the penalty 
    quadratization of Eq.~\eqref{eq:general_quadratization}. 
    Every vertex denotes a PUBO/QUBO configuration, edges represent 
    bit-flip neighbours. $\mathbf{x}_a$ and $\mathbf{x}_b$ 
    have different optimal $\mathbf{y}_{a}$ and $\mathbf{y}_b$, while 
    $\mathbf{y}_b$ is optimal for both $\mathbf{x}_b$ and $\mathbf{x}_c$.
    An unstable point $\mathbf{x}_c$ in PUBO can become a saddle in QUBO.}
    \label{fig:pubo_v_qubo_landscape}
    \vspace{-1em}
\end{figure}

\begin{figure}[!h]
    \vspace{-1em}
    \centering
    \includegraphics[width=0.7\columnwidth]{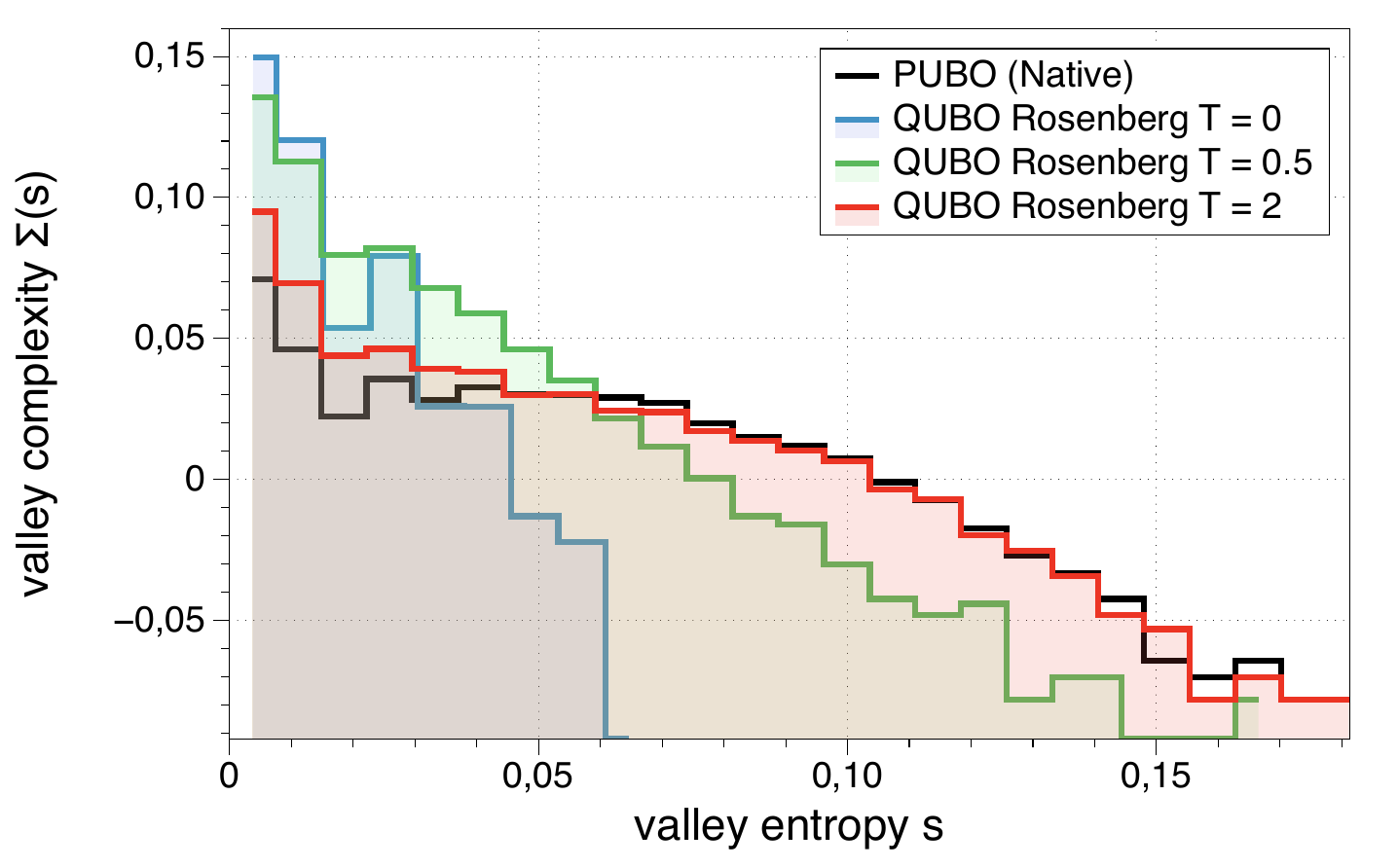}
    \caption{Sampled histogram of local minima valley complexity $\Sigma(s)$ vs valley entropy $s$ for the PUBO and Rosenberg QUBO landscapes, averaged over 100 
    SATLIB instances \textit{uf50-[901,1000]}. 
    Penalty hyperparameter for QUBO is $P = 0.5$.
    $T$ is temperature of QUBO connectivity method.}
    \label{fig:pubo_v_qubo_histogram}
     \vspace{-1.5ex}
\end{figure}

The degradation of the QUBO version of a higher-order problem is attributed to the substitution of original variables $x$ and augmenting the landscape with auxiliary variables $y$. This process not only increases exponentially the search space (Fig.~\ref{fig:qubobspubosearchspace}a), but also the new landscape is not faithful to the native space (see Fig.~\ref{fig:qubobspubosearchspace}b), whereby reducing QUBO energy does not perfectly correlate to solving more SAT clauses. 
An intrinsic property of the 3-SAT problem is 
the degeneracy of the configuration space, that it is possible to 
group configurations into large valleys based on their connectivity with each other
by zero-energy-change bit-flips (e.g. Fig.~\ref{fig:pubo_v_qubo_landscape} left), an HNN solver will diffuse through the optimization \textit{landscape} until a downward path is found, or a barrier at the valley border is overcome. When mapped to a QUBO problem $g(\mathbf{x}, \mathbf{y})$ with the Rosenberg penalty, 
\begin{equation}
    f(\mathbf{x}) = \min_{\mathbf{y}}g(\mathbf{x}, \mathbf{y})\,,
    \label{eq:general_quadratization}
\end{equation}
the QUBO landscape increases the ruggedness (e.g. Fig.~\ref{fig:pubo_v_qubo_landscape} right)
of the original PUBO manifold
due to auxiliary variables $\mathbf{y}$ for each fixed configuration $\mathbf{x}$.
To numerically estimate degenerate properties in PUBO/QUBO landscapes, we employ the Generalized Wang-Landau algorithm \cite{barbu2020} to sample the lowest 200 degenerate valleys for 100 instances from SATLIB \textit{(uf50-[901,1000])}. Following sampling, we discard valleys with zero-barrier exits (saddle points). For each remaining local minimum in a sampled PUBO degenerate valley, denoted as $\{\mathbf{x}\}^i$, we assess the potential for single bit-flip moves $\mathbf{x}_a^i \to \mathbf{x}_b^i$ in the QUBO space 
(in PUBO there is a zero barrier $E_{\mathrm{PUBO}}(\mathbf{x}_a^i)= E_{\mathrm{PUBO}}(\mathbf{x}_b^i)$) through the number $N_{\mathbf{y}}$ of combinations of auxiliary variables $\mathbf{y}$ satisfying $E_{\mathrm{QUBO}}(\mathbf{x}_a^i, \mathbf{y}) = E_{\mathrm{PUBO}}(\mathbf{x}_a^i)$. If a barrier $\Delta E(\mathbf{y}) = E_{\mathrm{QUBO}}(\mathbf{x}_b^i, \mathbf{y}) - E_{\mathrm{QUBO}}(\mathbf{x}_a^i, \mathbf{y})$ is overcome with probability $p = \frac{1}{N_{\mathbf{y}}}\sum_{\mathbf{y}}\exp{\left(-\Delta E(\mathbf{y})/T\right)}$, then we consider $\mathbf{x}_a$ and $\mathbf{x}_b$ connected in QUBO. Based on the resulting QUBO connectivity, we count the 
QUBO valley entropies $s$, i.e. the number of configurations $\exp{\{Ns\}}$ within a valley, 
and complexities $\Sigma(s)$, i.e. the number ($\exp{\{N\Sigma(s)\}}$) of valleys of size $s$, at different temperatures (see Fig.~\ref{fig:pubo_v_qubo_histogram}).

At small temperatures the QUBO landscape is very rugged, i.e. has more disconnected 
local minima valleys compared to PUBO,
which results in higher rejection of local moves and thus drastic performance slow-down. 
At higher temperature the QUBO energy barriers can of course be overcome and the entropic barriers 
of the PUBO landscape are restored. However, low-temperature performance determines convergence to solutions. The above results reinforce both quantitatively and conceptually the challenges of a quadratic-limited solver.

\section{28nm QUBO and PUBO Hardware Designs}\label{sec:hwmodels}
For quantitative analysis and comparison, we designed the needed hardware blocks for both QUBO and PUBO memristor-based solvers.  The hardware must offer full flexibility for current and future algorithm development, yet provide realistic performance metrics. Circuit elements at the device level were designed, optimized and simulated with respect to energy consumption and time requirements based on a TSMC 28nm CMOS technology. The central block is the VMM, which is built as a memristor-based CIM architecture \cite{Li2018}. The needed periphery includes driver circuits to activate word lines, current-voltage converters (transimpedance amplifiers, TIAs\cite{zimmermann2012}), DACs \cite{Karadimas2008} for additive noise signals in annealing, comparators \cite{Qadasi2018}, digital  random number generators (PRNGS) \cite{Marsaglia2003} and custom 1-out-of-n encoders (n/2-out-of-n encoders) for updating state vectors. 
\begin{figure*}[tbp]
	\centering
	\includegraphics[width=0.95\textwidth]{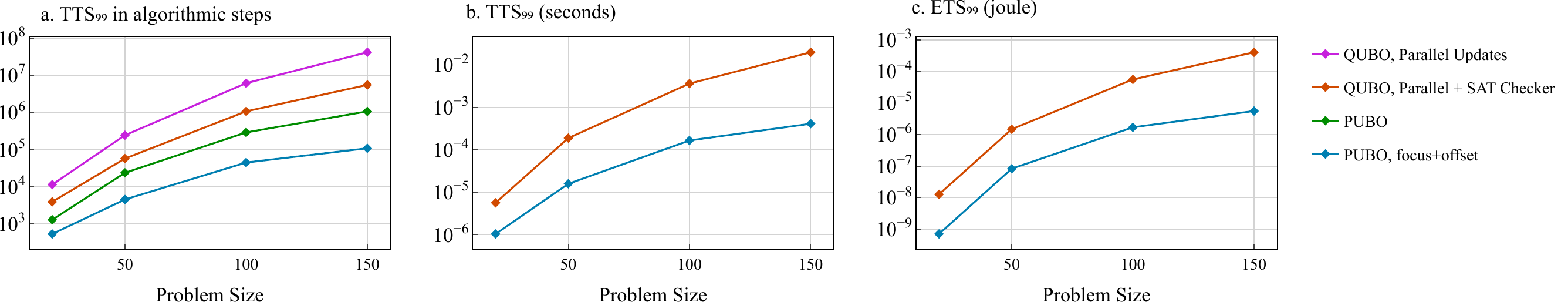}
	\caption{(Lower is better) a. TTS in algorithmic steps. b. True TTS (seconds) after circuit layout. c. ETS (Joules) for 0.99 success rate for 80 instances at each size. Curves are median values for each size.}
	\label{fig:ttscycle}
    \vspace{-2em}
\end{figure*}

\begin{figure}[!h]
    \centering
    \includegraphics[width=0.8\columnwidth]{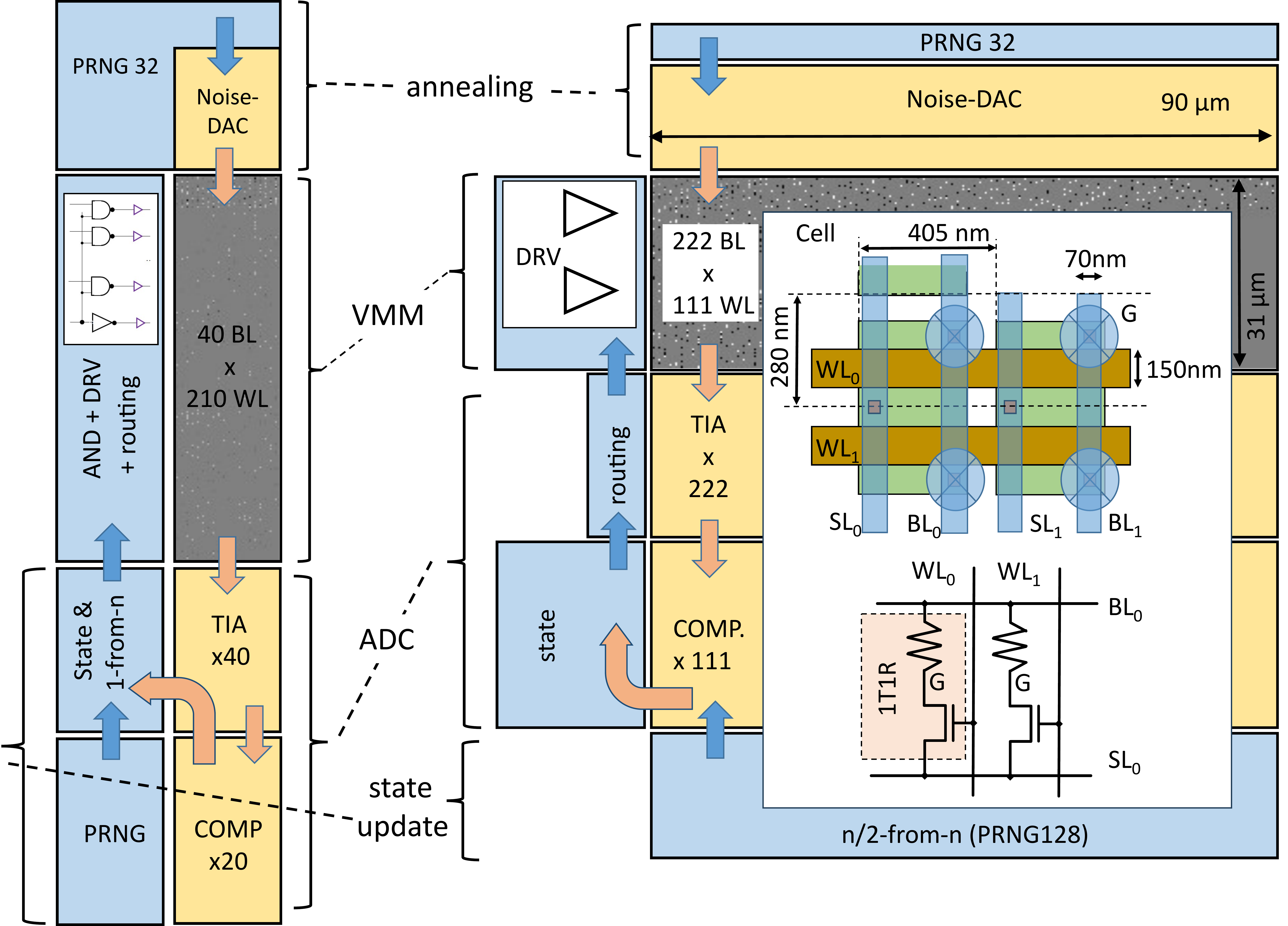}
    \caption{Floor plan of PUBO (left) and QUBO (right) for 20-variable 3-SAT problems with true-scaled circuit blocks based on 28 nm CMOS. Areas colored gray represent weight matrices, areas colored blue represent digital components, and areas colored ocher are analog/mixed-signal components.}
    \label{fig:floorplan}
    \vspace{-2ex}
\end{figure}

Circuit performance is expected to be dominated by an extended interconnect, thus it is first necessary to have an area planning of the individual components, after which an estimate of expected interconnect lengths of individual modules can be made. The interconnect lengths are then converted into effective capacitances and integrated into the Spectre circuit simulation model. From simulation, realistic values for delays and switching energies are obtained. 

\begin{figure}[!ht]
    \centering
    \includegraphics[width=1.0\columnwidth]{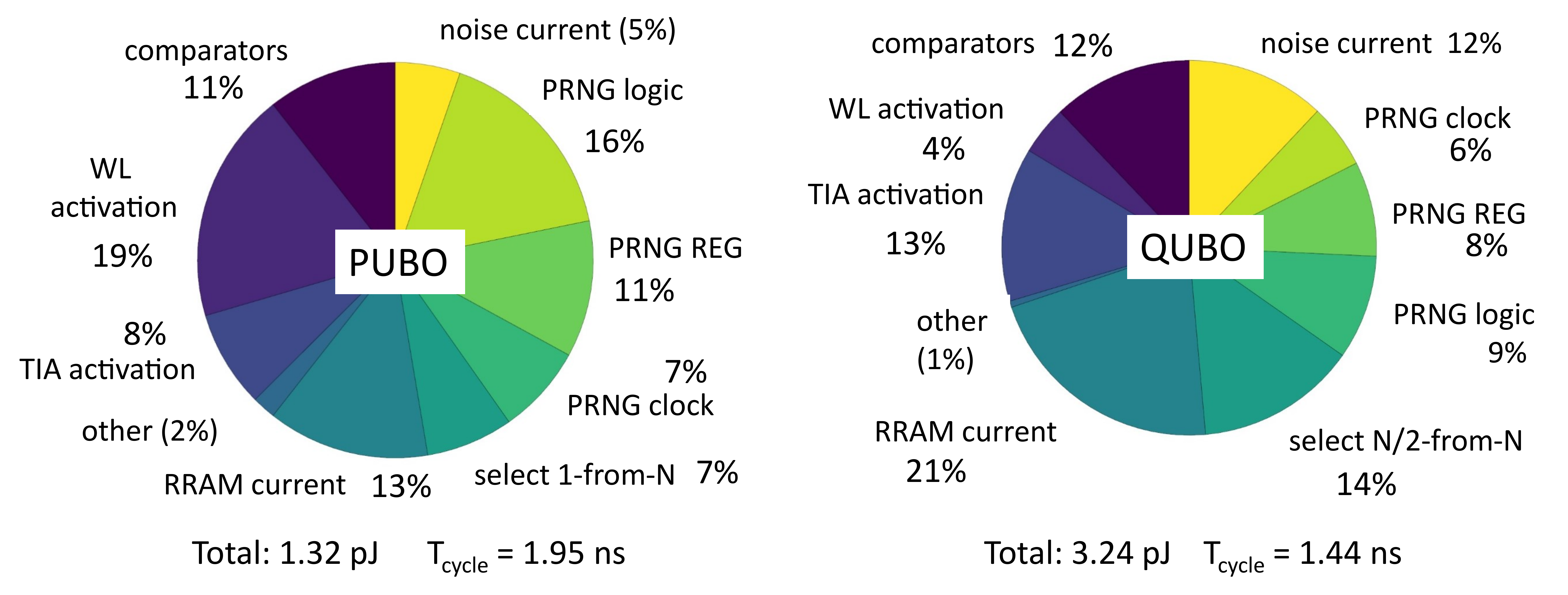}
    \caption{Energy total and breakdown for the various components in the PUBO (left) and QUBO (right) solvers.}
    \label{fig:piechart}
    \vspace{-2ex}
\end{figure}

Fig.~\ref{fig:floorplan} shows a designed floor plan of true-scale PUBO/QUBO implementations for 20-variable 3-SAT problems. Transient simulations of realistic state vector sequences provided numbers for aggregated power and timing numbers which were re-scaled to average energy contributions per cycle. Fig.~\ref{fig:piechart} shows the corresponding distribution of the energy contributions to the individual sub-operations.

\begin{figure}[!h]
    \vspace{-1em}
    \centering
    \includegraphics[width=0.9\columnwidth]{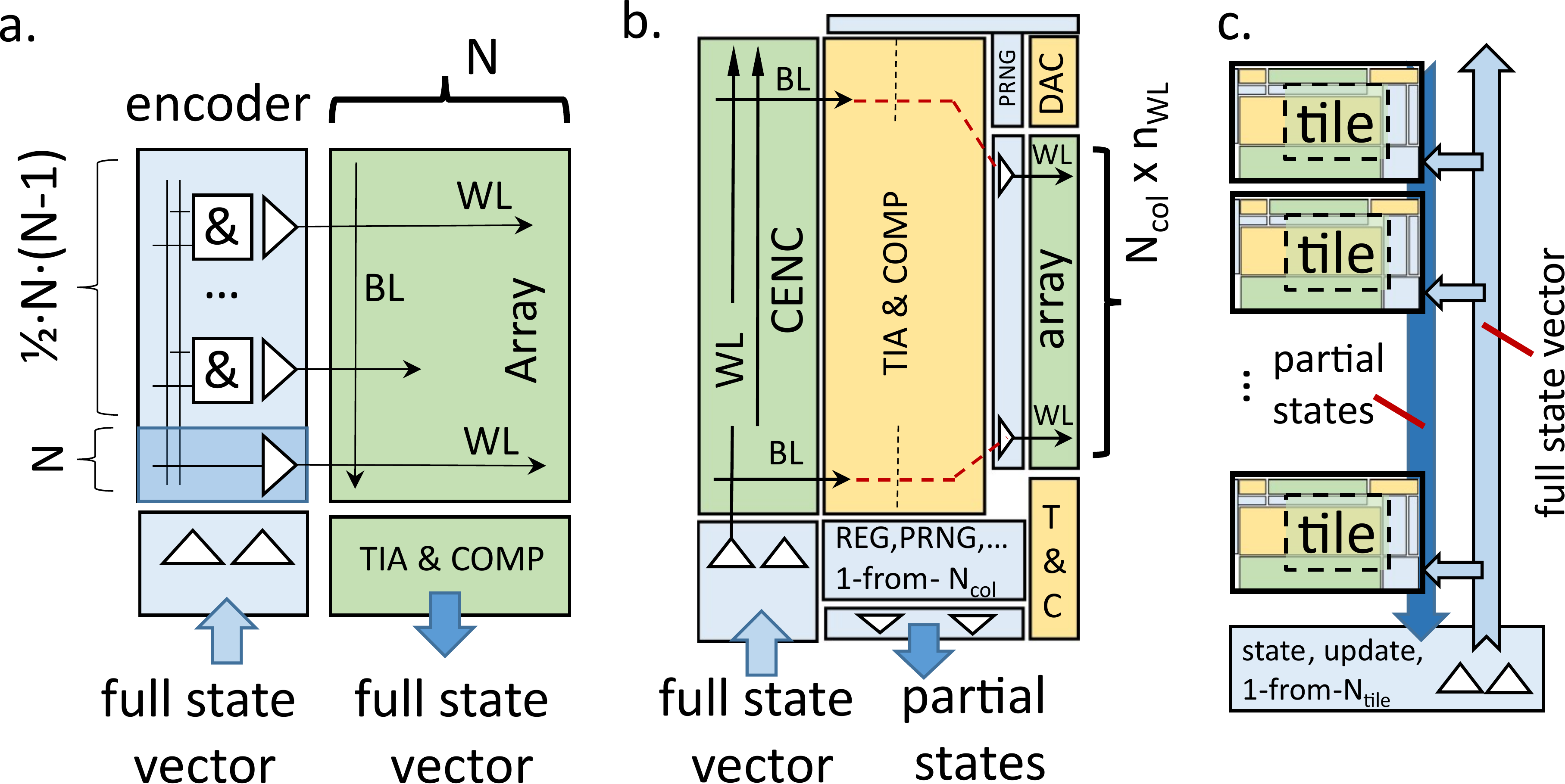}
    \caption{a. Conventional array periphery for PUBO, b. Configurable encoder (CENC) replacing the AND-encoder. True scale layout for $N=150$ and $N_{col}=19$, c. Tile architecture and interconnect.}
    \label{fig:tilingcam}
     \vspace{-1.5ex}
\end{figure}

Scaling PUBO/QUBO to problems with larger numbers of variables poses challenges for the hardware implementation. First, the number of word lines (PUBO) grows with $\sim \frac{1}{2} \text{N}(\text{N}-1)$ which leads to highly asymmetric, elongated layouts for $\text{N} \gg 20$. The gradient matrix (for well-randomized problems) contains many rows in which weights are zero.
With these aspects in mind, we developed a tiling architecture to support scaling. Instead of encoding all possible pairwise products of variables (see Fig.~\ref{fig:tilingcam}a), the encoder array is replaced by a memristor-based Configurable Encoder (CENC) (see Fig.~\ref{fig:tilingcam}). The CENC receives the complete state vector for input. For each horizontal bit line of the CENC block, a word line is now provided in the gradient array. The CENC array is programmed so that only active lines of the gradient matrix (associated to specific state bit combinations) are represented by corresponding patterns in the CENC.

The circuit components designed for the smaller 20-var implementation are utilized in the fully scalable CENC circuit. For a rectangular, fully filled layout of a tile, n\textsubscript{WL} = 400 with variable number of CENC inputs proves to be favorable. Each tile can thus accommodate sub-problems of larger problems.
For given 3-SAT problems from the SATLIB, an upper bound of possible state variables is obtained whose gradient can be fully mapped with a restricted and fixed number of word lines. In particular, for n\textsubscript{WL} = 400, N\textsubscript{col} = 19 state variables are obtained. An N-variable problem is thus mapped to $\text{N}_\text{tile} = \lceil \text{N}/\text{N}_\text{col}\rceil $ tiles (see Fig.~\ref{fig:tilingcam}c). A separate unit holds the state vector and the update logic, distributes the state vector as well as control signals to the tiles and collects the partial state vectors and other control signals from the tiles. Again, average energy and timing values are obtained from circuit simulation of the tiled architecture and translated into true TTS and ETS values (see Fig.~\ref{fig:ttscycle}(b and c)).

The quantitative hardware designs of QUBO and PUBO implementations are now combined with the algorithmic explorations of Section~\ref{sec:algcomp}.  We see (Fig.~\ref{fig:piechart}) that PUBO cycles consume $2.45\times$ less energy, but take $1.35\times$ longer than QUBO.  However, Section~\ref{sec:algcomp} showed a significant algorithmic advantage for PUBO, reducing the needed cycles to converge to solutions.  This PUBO advantage scales with problem size (Fig.~\ref{fig:ttscycle}a.), and our hardware analysis here quantifies the circuit costs associated with the problem scaling.  We thus combine both the hardware and algorithmic analysis in Figs.~\ref{fig:ttscycle}b and c to yield time-to-solution and energy-to-solution numbers.  Our analsyis shows a substantial and growing advantage of PUBO, with $48\times$ and $72\times$ improved speed and energy, respectively, over QUBO at the largest problem sizes.  This performance gap is expected to continue for larger problems, which is future work.
\vspace{-2ex}

\section{Conclusion}
In this work we developed a memristor-based hardware implementation of a higher-order HNN (i.e. PUBO) showing significantly reduced time and energy consumption compared to more standard quadratic IM (QUBO). We separately quantified the gains from both the hardware and algorithmic aspects. We showed operating in the higher-order search space reduced variables (exponentially smaller search space), provides a more faithful energy landscape to the original problem landscape, and can be smoother for the solver to explore.

\section{Acknowledgment}
We gratefully acknowledge computing time on the supercomputer JURECA \cite{jureca} at Forschungszentrum Jülich under grant no. `optimization'. We also are gratfully acknowledge the generous funding of this work with is under NEUROTEC II (Verbundkoordinator / Förderkennzeichen: Forschungszentrum Jülich: 16ME0398K) by the Bundesministerium für Bildung und Forschung. This project was also under the DARPA QuICC project under contract no. FA8650-23-3-7313.

\bibliographystyle{IEEEtran}
\bibliography{main}
\end{document}